\documentclass[aps,prb,groupedaddress,floatfix]{revtex4}

\usepackage{graphicx}
\usepackage{subfigure}
\usepackage{epsfig}
\usepackage{amsmath}
\usepackage{bbm}

\def\be{\begin{equation}}
\def\ee{\end{equation}}
\def\bea{\begin{eqnarray}}
\def\eea{\end{eqnarray}}

\begin{document}


\title{Critical currents in graphene Josephson junctions}


\author{J. Gonz{\'a}lez$^1$ and E. Perfetto$^{2}$}
\affiliation{$^{1}$Instituto de Estructura de la Materia.
        Consejo Superior de Investigaciones Cient{\'\i}ficas.
        Serrano 123, 28006 Madrid. Spain.\\
             $^{2}$Consorzio Nazionale Interuniversitario per le Scienze
Fisiche della Materia, Universit\`a di Roma Tor Vergata, Via della
Ricerca Scientifica 1, 00133 Roma, Italy.}

\date{\today}

\begin{abstract}
We study the superconducting correlations induced in graphene when it is 
placed between two superconductors, focusing in particular on the supercurrents 
supported by the 2D system. For this purpose we make use of a formalism placing 
the emphasis on the many-body aspects of the problem, with the aim of
investigating the dependence of the critical currents on relevant variables
like the distance $L$ between the superconducting contacts, the temperature,
and the doping level. Thus we show that, despite the vanishing density of 
states at the Fermi level in undoped graphene, supercurrents may exist at zero 
temperature with a natural $1/L^3 $ dependence at large $L$. When temperature 
effects are taken into account, the supercurrents are further suppressed beyond 
the thermal length $L_T $ ($\sim v_F / k_B T $, in terms of the Fermi velocity 
$v_F $ of graphene), entering a regime where the decay is given by a 
$1/L^5 $ dependence. On the other hand, the supercurrents can be enhanced upon
doping, as the Fermi level is shifted by a chemical potential $\mu $ from 
the charge neutrality point. This introduces a new crossover length 
$L^* \sim v_F / \mu $, at which the effects of the finite charge density start 
being felt, marking the transition from the short-distance $1/L^3 $ behavior to 
a softer $1/L^2 $ decay of the supercurrents at large $L$. It turns out that
the decay of the critical currents is given in general by a power-law behavior, 
which can be seen as a consequence of the perfect scaling of the Dirac theory 
applied to the low-energy description of graphene.

\end{abstract}

\maketitle

\section{Introduction}

Since the discovery of single atomic layers of carbon in 2004 \cite{novo}, 
this new two-dimensional (2D) material (so-called graphene) has attracted 
a lot of attention\cite{att}. 
From the experimental point of view, the 2D carbon sheets have 
shown a number of remarkable electronic properties. Thus, there has been 
evidence that graphene may have a finite lower bound ($4 e^2 /h$) in the 
conductivity at the charge neutrality point\cite{geim,kim}. 
Furthermore, an anomalous integer
quantum Hall effect has been measured in the 2D system with plateaus at 
odd-integer values of the quantum of conductance\cite{geim,kim}. 
The absence of weak localization effects\cite{mor} 
has also pointed at the unconventional effects that 
impurities and in general disorder may produce in the graphene sheet.

Most of the remarkable transport properties of graphene have to do with 
its particular band structure at low energies. The undoped system has a 
finite number of Fermi points, placed at the corners of the hexagonal 
Brillouin zone. Only two of such points can be taken as independent, with 
quasiparticle excitations which have conical dispersion above and below the 
Fermi level\cite{wall}. 
This explains that the low-energy electronic states of graphene may be 
accommodated into two two-component spinor fields, governed by a Dirac 
hamiltonian which leads to a dispersion relation 
$\varepsilon ({\bf k}) = \pm v_F |{\bf k}|$. The electronic system displays 
hence a relativistic-like invariance at low energies, which is at the origin 
of the finite lower bound in the conductivity\cite{paco,kat,twor,mac}, the
anomalous integer Hall effect\cite{paco,ando,gus}, and the absence of
backscattering in the presence of long-range scatterers\cite{suzu}. 
Other exotic effects relying on the Dirac theory have been proposed, like 
the selective transmission of electrons through a $n$-$p$ junction\cite{falk}
or the specular Andreev reflection at a graphene-superconductor 
interface\cite{been}.

Recently, the properties of graphene have been also investigated when the 
material is placed between superconducting contacts. Thus, in the experiment 
reported in Ref. \onlinecite{delft}, it has been possible to measure
supercurrents in graphene by attaching wide superconducting electrodes with a 
spatial separation of $\approx 0.5 \; \mu$m. In another experiment, reported 
in Ref. \onlinecite{orsay}, a quite different geometry has been investigated by 
placing thin electrodes across a large 2D sample, with a minimum separation 
between the tips of $\approx 2.5 \; \mu$m. In this case, the evidence 
of the superconducting correlations in graphene has been obtained in the form 
of Andreev reflection peaks in the $I$-$V$ curves, as well as in the abrupt drop 
of the resistance at a temperature of $\approx 1$ K, below the 
critical temperature ($\approx 4$ K) of the superconducting electrodes.
Moreover, supercurrents have been also measured in the experiment reported in
Ref. \onlinecite{brd}, where their development may have been favored by 
the large aspect ratio ($\sim 10$) between the width of the junction and the 
lead separation (of the order of a few hundreds of nanometers).

It is therefore pertinent to study the way in which the superconducting 
correlations are induced in graphene when it is placed between two 
superconductors, and how such correlations may depend on the geometry of the 
experimental setup. In this paper we are going to address this issue, focusing 
in particular on the supercurrents supported by the graphene 
sheet. We will be using a formalism placing the emphasis on the many-body 
aspects of the problem. This will allow us to clarify a number of questions, 
regarding the dependence of the critical currents on relevant variables like 
the distance between the superconducting contacts, the temperature, and the 
doping level of the graphene sample. In this respect, our approach can be seen 
as complementary to that of Ref. \onlinecite{tit}, 
where the Josephson effect has been studied in terms of Andreev reflection at 
superconducting contacts, concentrating on junctions with relatively short 
distance between the electrodes. We will be dealing with a framework where 
the tunneling and propagation of the Cooper pairs in graphene play the central 
role, placing in principle no restriction on the separation that may exist 
between superconducting contacts.

The content of this paper is distributed as follows. We will set up in section
II the formalism needed to describe the tunneling and propagation of Cooper 
pairs in graphene. This will be applied to the computation of the critical 
currents in section III, where we will also discuss the different regimes 
depending on the interplay between the temperature and the distance between 
superconducting contacts. Section IV will be devoted to extend our analysis
to the case of finite doping, showing the enhancement experienced then by the 
supercurrents. Finally, we will summarize our results and draw our conclusions
in section V.

\section{Model of graphene Josephson junction}

Our purpose is to build a model that incorporates the low-energy 
properties of electron quasiparticles in graphene as well as the 
tunneling of electrons from graphene to the superconducting
electrodes and vice versa. We take into account in particular that,
below an energy scale of $\sim 1$ eV, the electron dispersion relation
has a conical shape, with a dependence of the energy $\varepsilon $ on 
momentum ${\bf k}$ given by 
$\varepsilon ({\bf k}) \approx \pm v_F |{\bf k}|$ \cite{wall}. We have to 
bear in mind that the 2D system has actually two independent
Fermi points supporting such a conical dispersion, at opposite corners
$K, -K$ of the hexagonal Brillouin zone. The dynamics of the quasiparticles 
in graphene can be therefore described in terms of a couple of two-component
Dirac spinors $\Psi^{(a)} $, $a = 1, 2$, with a hamiltonian\cite{mele,nos}
\begin{equation}
H_0   =   v_F \int d^2 r \; \Psi^{(a) \dagger}_{\sigma} (\mathbf{r})
  \:  \mbox{\boldmath $\sigma$}^{(a)}\cdot\mbox{\boldmath $\partial$} \:
              \Psi^{(a)}_{\sigma} (\mathbf{r})
\label{h0}
\end{equation}
where $\{ \mbox{\boldmath $\sigma$}^{(a)} \}$ are two different suitable
sets of Pauli matrices\cite{ando}  (we use units such that $\hbar = 1$). 
In the above expression, the label of the spinor
components is omitted for simplicity, and a sum is taken implicitly 
over the spin index $\sigma $ as well as over the index $a$ running over 
the two different low-energy valleys of the dispersion. 

The above hamiltonian has to be then complemented with a term accounting 
for the tunneling of electrons from the graphene side to the superconducting 
electrodes and vice versa. In this respect, we are going to assume that 
the tunneling takes place with equal amplitude for the two sublattices of
the graphene honeycomb lattice. This kind of junction may be realized in
cases where the contacts between graphene and the superconductors preserve
the structure of the graphene lattice. From a technical point of view, such
condition implies that the different spinor components and the different
low-energy valleys couple with equal amplitude to the superconductors. 
By denoting the electron fields in the respective superconducting electrodes 
by $\Psi_{S1}$ and $\Psi_{S2}$, we may write the tunneling hamiltonian
for contacts along the coordinates $x_1 = 0$ and $x_2 = L$ as
\begin{equation}
H_t  =   \sum_{j=1,2} t  \int_0^W dy \;
        \Psi^{(a) \dagger}_{\sigma} (x_j,y)  \Psi_{Sj,\sigma} (x_j,y)
                           +  {\rm h.c.}
\label{tun}
\end{equation}
where the parameter $t$ represents the tunneling amplitude. 
We stress at this point that, while the contacts 
have a width given by $W$ in Eq. (\ref{tun}), the extension of the 
graphene layer along the transverse $y$ direction is not constrained by this 
parameter in our model. Thus, our description will apply in general to
2D graphene samples, with dimensions in both the transverse and the 
longitudinal direction much larger than the contacts introduced by the 
superconducting electrodes. 

The properties of the superconducting electrodes have to be also 
incorporated in the model of the Josephson junction. For the description 
of the supercurrents, it will be enough to specify the normal density of 
states $\rho $ and the order parameter $\Delta $ in the superconducting 
state. We recall that a supercurrent arises in general from a gradient in
the phase of the order parameter in a superconductor. In the case of a 
Josephson junction, the supercurrent is produced by a mismatch in the 
phases $\chi_1 $ and $\chi_2 $ of the respective order parameters in the
superconducting electrodes. The Josephson current $I_s $ is actually
given by the derivative of the free energy with respect to the variable
$\chi = \chi_1 - \chi_2 $, and it can be therefore expressed as
\begin{equation}
I_s = 2e \frac{\partial }{\partial \chi}
k_B T \: \log \left( {\rm Tr} \: e^{-H/k_B T} \right) 
\label{joseph}
\end{equation}
where $T$ is the temperature and $H$ stands for the full 
hamiltonian of the model.

\begin{figure}[h]
\begin{center}
\epsfxsize 6cm \epsfbox{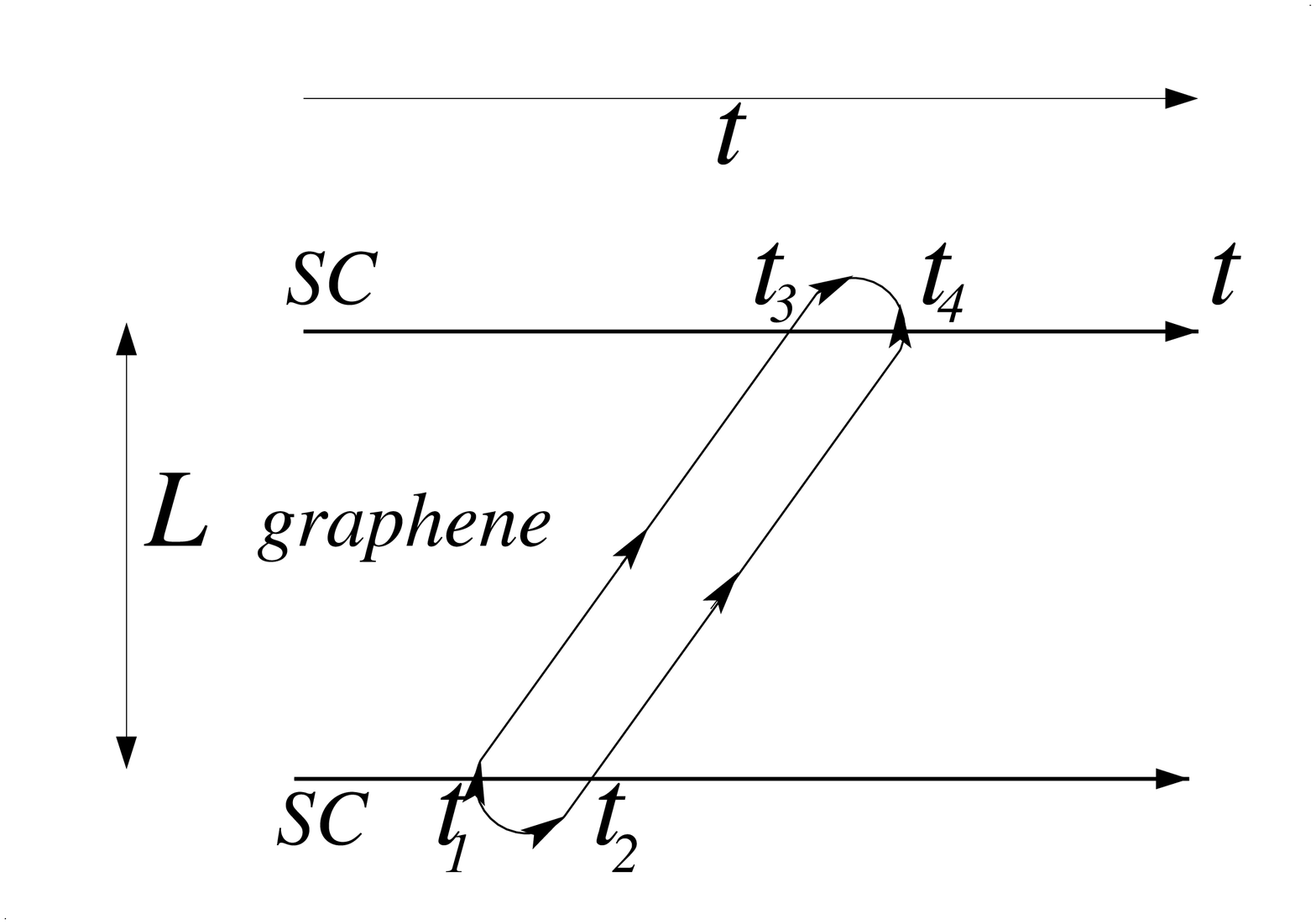}
\end{center}
\caption{Schematic representation of the propagation of Cooper pairs in
graphene between two superconductors (SC).}
\label{tun1}
\end{figure}

In order to compute the Josephson current from Eq. (\ref{joseph}), we
will resort to a perturbative expansion in the tunneling 
amplitude $t$. The structure of the dominant contributions may be however very 
different depending on the actual geometry of the Josephson junction\cite{faz}. 
In cases where the distance $L $ between the contacts is much smaller than 
the superconducting coherence length $\xi $, the supercurrents are built from 
processes with independent tunneling and uncorrelated 
propagation in graphene of the electrons of a Cooper pair. On the other hand, 
when $L$ is much larger than $\xi $, the behavior is governed by the fast 
tunneling and subsequent propagation of the Cooper pair in graphene, as shown 
schematically in Fig. \ref{tun1}. This 
situation corresponds to the case where the time of propagation between the
contacts is much larger than $1/|\Delta |$. Under the assumption of a large
$|\Delta |$, the relevant properties of the superconductors may be encoded 
in the statistical average  
\begin{equation}
 \langle \Psi_{Sj,\sigma} (x_j,y;-i\tau_1) 
     \Psi_{Sj,-\sigma} (x_j,y;-i\tau_2)  \rangle  \approx 
     e^{i \chi_j} \rho \: \delta (\tau_1 - \tau_2 )
\label{cond}
\end{equation}
where the operators are ordered with respect to imaginary time $\tau $.

From inspection of the expansion of 
the r.h.s. in Eq. (\ref{joseph}) in powers of the tunneling amplitude,
we observe that the first nonvanishing contribution to $I_s $ appears to 
fourth order in $t$, from a statistical average of operators 
participating of the condensates of the two superconductors.
The expression of the maximum supercurrent $I_c$ (critical current) is
worked out at that perturbative level in the Appendix, focusing on the regime 
corresponding to $L \gg \xi$. After factoring out
the relative tunnel conductances at the contacts (given in each case by the 
dimensionless quantity $\rho t^2 W /v_F $), we end up with an expression
for the behavior of the critical current intrinsic to the 2D graphene layer: 
\begin{eqnarray}
I_c^{(2D)} (T) & \approx &  2e v_F^2 \int_0^W dy_1 \int_0^W dy_2
     \int_0^{1/k_B T}  d \tau    \:                     
      \langle \Psi^{(a) \dagger}_{\uparrow} (0,y_1;0)
            \Psi^{(-a) \dagger}_{\downarrow} (0,y_1;0)   
              \Psi^{(b)}_{\uparrow} (L,y_2; -i\tau )
             \Psi^{(-b)}_{\downarrow} (L,y_2; -i\tau )   \rangle
\label{crit}
\end{eqnarray}
We observe from (\ref{crit}) that the propagator of the Cooper pairs
evaluated over a distance $L$ plays the central role in the determination
of the supercurrents. We will study in what follows the behavior
of this propagator depending on the distance $L$, the temperature, and the 
doping level.

\section{Supercurrents at finite temperature}

We analyze first the behavior of the supercurrents in graphene when the 
system is undoped, but is placed at a nonvanishing temperature $T$. 
The expectation values in the above formulas have to be understood then as 
statistical averages at that finite temperature. The building block for all 
the calculations is the electron propagator
\begin{equation}
G^{(a)} ({\bf r}, t) = 
 -i \langle T \Psi^{(a)}_{\sigma} (0,0) 
      \Psi^{(a) \dagger}_{\sigma} ({\bf r}, t) \rangle
\end{equation}
This is given in graphene by the propagator for Dirac fermions, 
in correspondence with the hamiltonian (\ref{h0}). In the 
many-body theory at temperature $T \neq 0$, the imaginary part of that object 
gets a specific term to account for the thermal effects. The full
expression of the Dirac propagator becomes in momentum space\cite{dj}
\begin{equation}
G^{(a)} ({\bf p}, \omega_p) = \frac{\omega_p + \mbox{\boldmath $\sigma$}^{(a)} 
  \cdot {\bf p} }{ \omega_p^2 - {\bf p}^2 + i \epsilon }
 + i 2\pi ( \omega_p + \mbox{\boldmath $\sigma$}^{(a)} \cdot {\bf p} )
   \delta ( -\omega_p^2 + {\bf p}^2 ) \frac{1}{1 + e^{|\omega_p |/k_B T }}
\label{diract}
\end{equation}


The Cooper-pair propagator in (\ref{crit}) can be computed from the 
convolution of two Dirac propagators, bearing in mind that they correspond to 
fields at opposite valleys of the graphene dispersion. In doing this operation, 
we will have to be also consistent with our assumption that the tunneling at 
the superconducting contacts is the same for the two sublattices of the 
graphene lattice. This means that, when taking the average for the Cooper-pair 
propagator, we will also take a trace in spinor space over the states of the 
Cooper pairs in sublattice $A$, given by 
$\Psi^{(a)}_{A,\uparrow }({\bf k} + {\bf q}) \Psi^{(-a)}_{A,\downarrow }(-{\bf q})$, 
and in sublattice $B$, given by
$\Psi^{(a)}_{B,\uparrow }({\bf k} + {\bf q}) \Psi^{(-a)}_{B,\downarrow }(-{\bf q})$.
The Cooper-pair propagator thus defined in momentum space, 
$D ({\bf k}, \omega )$, can be expressed as 
\begin{equation}
D ({\bf k}, \omega_k ) = i \: {\rm Tr } \int \frac{d \omega_q}{2 \pi} 
  \int \frac{d^2 q}{(2 \pi)^2} G^{(a)} ({\bf q}+{\bf k}, \omega_q + \omega_k)
  G^{(-a)} (-{\bf q}, -\omega_q )
\label{freeprop}
\end{equation}

A nice feature of the diagrammatics of the many-body theory at $T \neq 0$ is 
that the terms carrying the dependence on temperature do not need to be
regularized by means of a high energy cutoff. The contributions at $T = 0$, 
however, remain finite only when the integrals over the momenta are 
suitably cut off. In the present model, it is convenient to choose a method
of regularization of the integrals preserving the relativistic-like invariance
of the theory. For this purpose, we will adopt an analytic continuation in 
the number of space-time dimensions\cite{nos2}, that is, carrying out first 
the integrals at general dimension $D$, and then taking the limit 
$D \rightarrow 3$. To implement this procedure, we first collect the components 
of the momentum and the frequency to form 3D vectors, 
$q \equiv (v_F {\bf q}, \omega_q )$, $k \equiv (v_F {\bf k}, \omega_k )$. 
Next, we may rotate all the 3D vectors to Euclidean space by introducing 
imaginary frequencies, $\overline{\omega}_q = -i \omega_q $. One can easily
see that the expression of the propagator (\ref{freeprop}) at general 
dimension $D$ becomes
\begin{eqnarray}
 \left. D ({\bf k}, i \overline{\omega}_k) \right|_{T = 0}   & = &
\int_0^1 dx  \int \frac{d^D q}{(2 \pi )^D} \frac{2 q^2 - 2 k^2 x(1-x)}
                {\left( q^2 + k^2 x(1-x) \right)^2}           \nonumber    \\
 & = & \left(  \frac{1}{4 \pi^{3/2}} \Gamma \left(1 - \frac{D}{2} \right)  -  
       \frac{1}{2 \pi^{3/2}} \Gamma \left(2 - \frac{D}{2} \right)  \right)
           \int_0^1 dx \sqrt{k^2 x(1-x)}
\label{dimreg}
\end{eqnarray}
In the last passage we have made use of standard formulas in dimensional
regularization. Quite remarkably, the result turns out to be finite in the 
limit $D \rightarrow 3$. After reverting the rotation back to real frequency, 
we finally get 
\begin{equation}
 \left. D (\mathbf{k}, \omega) \right|_{T = 0} =
    -\frac{1}{8 v_F^2} \sqrt{v_F^2 \mathbf{k}^2 - \omega^2}
\label{zerot}
\end{equation}

The part of the Cooper-pair propagator depending on temperature can be 
computed by using the second term in (\ref{diract}) to make the convolution
(\ref{freeprop}). For our purposes, we can concentrate on the calculation
of the Cooper-pair propagator at zero frequency. By adding the result
(\ref{zerot}) to the temperature-dependent contribution, we get
\begin{equation}
D ({\bf k}, 0) = -\frac{1}{8 v_F} |{\bf k}| 
 - \frac{\log(2)}{\pi v_F^2} k_B T 
 + \frac{1}{2 \pi v_F} |{\bf k}| 
   \int_0^1 dx \frac{1}{\sqrt{1-x^2}} \frac{1}{1 + e^{x v_F |{\bf k}|/2k_B T}}
\label{propt}
\end{equation}

From the results (\ref{zerot}) and (\ref{propt}), we can already extract a
number of conclusions regarding the behavior of the supercurrents in long
graphene Josephson junctions. From Eq. (\ref{crit}), we can express the 
critical current for $L \gg W$ as 
\begin{equation}
I^{(2D)}_c (T)     \approx    2e v_F^2 W^2   \int_0^{\infty }
  \frac{dk}{2\pi } \; |{\bf k}| \; J_0 (|{\bf k}| L) D({\bf k}, 0)
      e^{- |{\bf k}| /k_c }
\label{2D}
\end{equation}
A short distance cutoff $k_c $ has been introduced to regularize the integral 
over the momentum. This is actually justified on physical grounds,
since the description of graphene as a continuum in terms of the Dirac theory
makes sense at distances above the nanometer scale. A sensible choice 
corresponds to $v_F k_c \sim 1$ eV. We will see that, at distances such that 
$L \gg k_c^{-1}$, the behavior of the critical current is in general not
sensitive to the actual value of the cutoff.

At $T = 0$, the dependence of the critical current on $L$ can be 
obtained from the Cooper-pair propagator (\ref{zerot}). Actually, we can 
derive an analytical expression for $I^{(2D)}_c (0)$ by computing the 
integral in (\ref{2D}):
\begin{eqnarray}
I^{(2D)}_c (0)  & \sim &  - e v_F W^2 \int_0^{\infty } dk \; |{\bf k}|^2 \;
               J_0 (|{\bf k}| L)    e^{- |{\bf k}| /k_c }             \\ 
     & =  &          e v_F W^2 
       \frac{k_c^3 (k_c^2 L^2 - 2)}{\sqrt{(k_c^2 L^2 + 1)^5}}         \\ 
\end{eqnarray}
From this result we check that, as expected, the behavior of the critical
current is not affected by the cutoff $k_c $ in the limit of large $L$.
In this regime we find
\begin{equation}
I^{(2D)}_c (0)    \sim    e v_F W^2 \frac{1}{L^3 }                      
\label{powerl}
\end{equation}
The strong power-law decay shown by (\ref{powerl}) can be understood actually 
as a reflection of the linear dependence on momentum 
of the quasiparticle energy, which dictates in turn the behavior
of the Cooper-pair propagator (\ref{zerot}) \cite{prb}.
We reach anyhow the interesting conclusion that, while graphene has a 
vanishing density of states at the Dirac point, it may still support
a nonvanishing supercurrent when the Fermi level is at that charge 
neutrality point.

\begin{figure}
\begin{center}
\epsfxsize 5cm \epsfbox{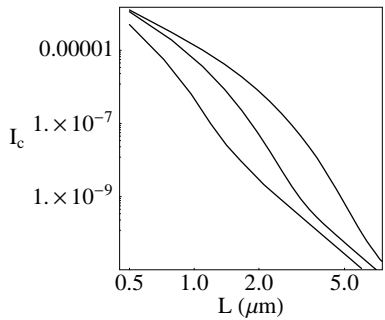}
\end{center}
\caption{Logarithmic plot of the critical current $I_c^{(2D)}$ (in units of 
$10^{-2} e v_F k_c \approx 1.2 \; \mu$A) as a function of the distance $L$, 
taking $W = 10^2 /k_c$ ($= 50$ nm). The three curves correpond, from top 
to bottom, to different values of the temperature $T = 2$ K, 4 K, and 8 K.}
\label{one}
\end{figure}

The inspection of the full propagator (\ref{propt}) also reveals that the
scaling is drastically modified when $k_B T \gg v_F |\bf{k}|$. Actually,
we can distinguish between a high-temperature and a low-temperature regime
of the Cooper-pair propagator, with quite different behaviors:
\begin{eqnarray}
D ({\bf k}, 0)  & \approx &  - \frac{1}{8 v_F} |{\bf k}|  \;\;\;\;\;
  \;\;\;  \;\;\; \;\;\;  \;\;\;
           {\rm if}  \;\;\;   k_B T  \ll  v_F |{\bf k}|                \\
        & \approx &  - \frac{\log (2)}{\pi v_F^2} k_B T 
    - \frac{1}{16 \pi }  \frac{|{\bf k}|^2}{k_B T}
          \;\;\;\;\;  {\rm if}  \;\;\;    k_B T \gg  v_F |{\bf k}| 
\label{scalingt}
\end{eqnarray}
The existence of this crossover in the momentum gives rise to an abrupt 
decay of the supercurrent beyond the thermal length $L_T = v_F /k_B T $.
This is illustrated in Fig. \ref{one}, where the critical current
$I^{(2D)}_c (T)$ is represented as a function of the distance $L$ at 
different temperatures. We observe for instance that, for a temperature 
of the order of $T \sim 1$ K, the scale of the crossover in $L$ is of the 
order of a few microns, in agreement with the expression of the 
thermal length.

From a physical point of view, it becomes clear that the 
Cooper pairs do not feel the thermal effects during their propagation 
when $L$ is shorter than the scale given by $L_T$, while they are
increasingly disrupted at distances larger than the thermal length.
At short distances such that $L \ll v_F /k_B T $, the decay of the 
critical current represented in Fig. \ref{one} follows a $1/L^3$ power-law,
in agreement with the above analysis at $T = 0$. However, beyond the 
crossover clearly identified in the three curves, we see that a different 
power-law behavior opens up at long distance $L \gg v_F /k_B T $. This 
regime can be analyzed by considering that, when $T$ is very large, the 
second term in the approximation (\ref{scalingt}) dictates the 
long-distance decay of the critical current. In this case we can compute
again analytically the integral in (\ref{2D}):
\begin{eqnarray}
I^{(2D)}_c (T)  & \sim &  - e v_F^2 W^2  
            \frac{1}{k_B T} \int_0^{\infty } dk \; |{\bf k}|^3 \;
               J_0 (|{\bf k}| L)    e^{- |{\bf k}| /k_c }             \\ 
     & =  &          e v_F W^2  \frac{v_F}{k_B T}  
       \frac{k_c^4 (9 k_c^2 L^2 - 6)}{\sqrt{(k_c^2 L^2 + 1)^7}}        
\end{eqnarray}
The leading contribution to the critical current becomes then for 
$L \gg v_F /k_B T $
\begin{equation}
I^{(2D)}_c (T)   \sim   e v_F W^2 \frac{v_F}{k_c k_B T} \frac{1}{L^5 }
\label{strong}
\end{equation}
The existence of this stronger power-law decay is manifest in the results 
of the numerical computation of the critical current represented in Fig.
\ref{one}, as it can be checked that the rightmost part of the lower curves 
in the plot corresponds with great accuracy to a power-law behavior 
with the exponent given by Eq. (\ref{strong}).

\begin{figure}
\begin{center}
\mbox{\epsfxsize 5cm \epsfbox{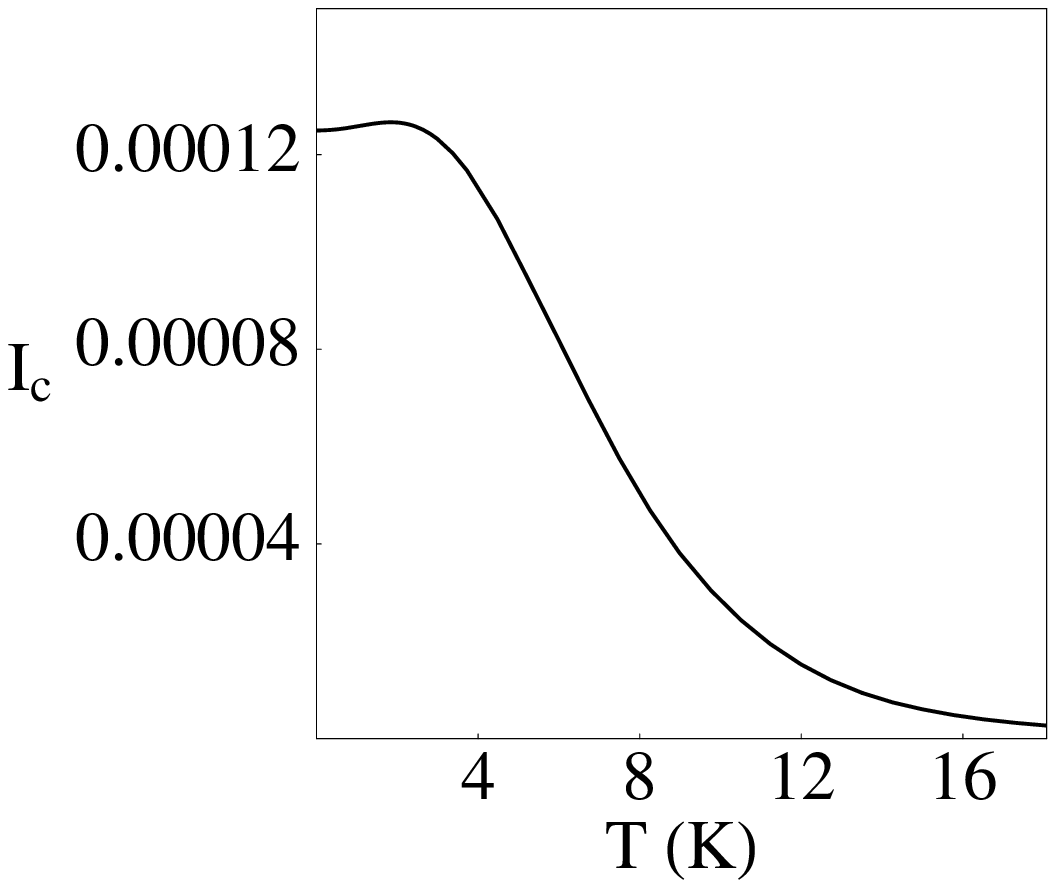} \hspace{0.45cm}
\epsfxsize 5cm \epsfbox{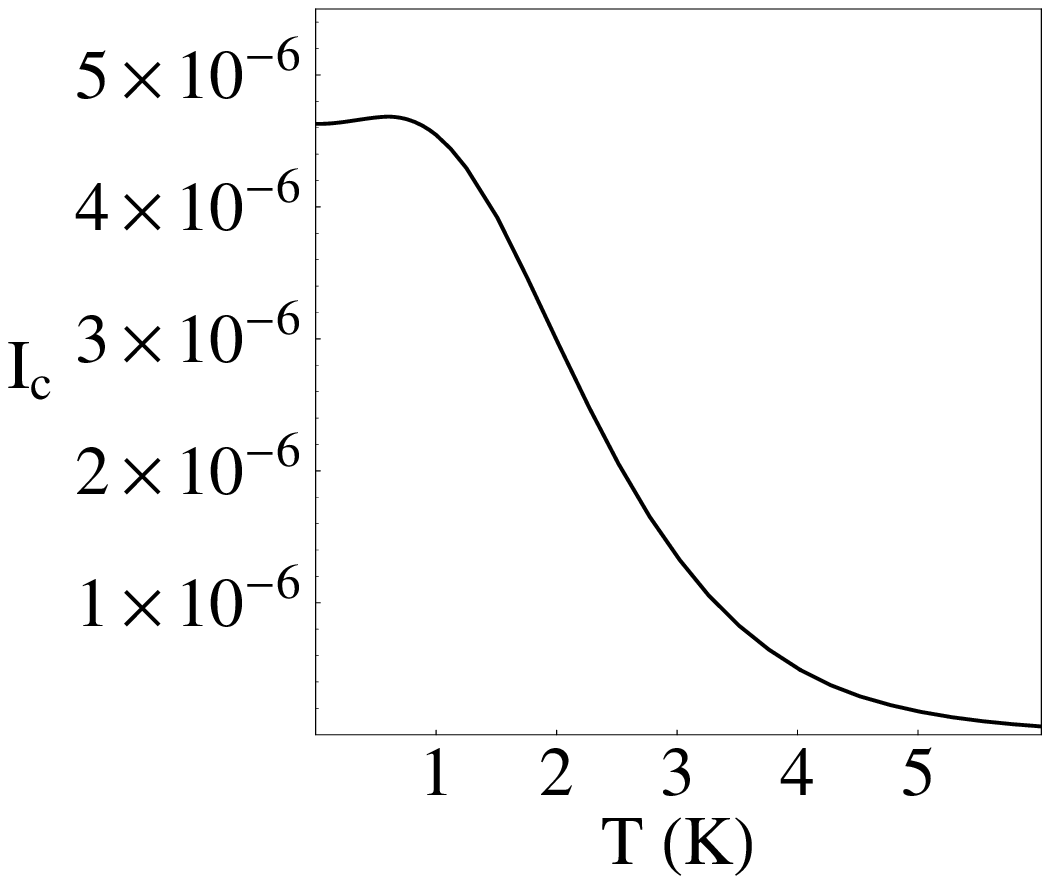}      \hspace{0.45cm}
\epsfxsize 5cm \epsfbox{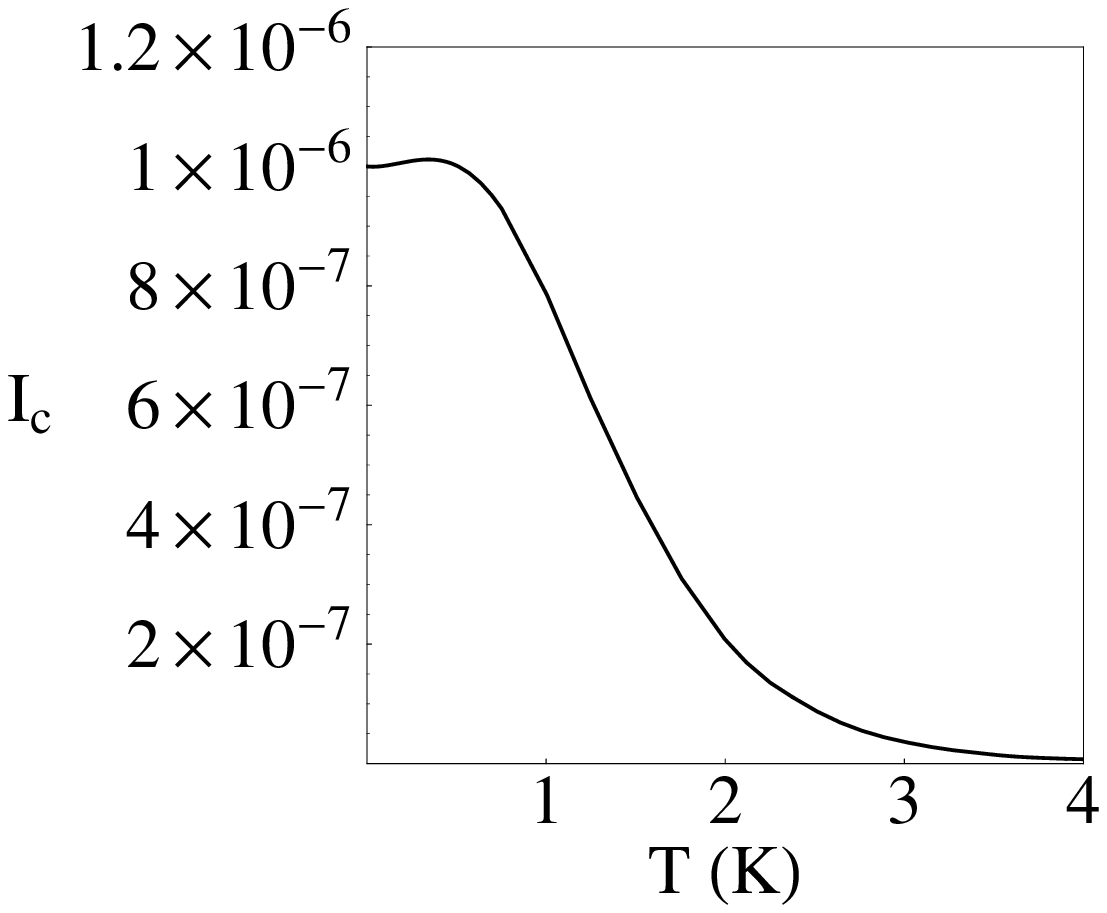}}\\
 \hspace{1.25cm}   (a) \hspace{5.05cm} (b)   \hspace{5.0cm} (c)
\end{center}
\caption{Plot of the critical current $I_c^{(2D)} (T)$ (in units of 
$10^{-2} e v_F k_c \approx 1.2 \; \mu$A) as a function of the temperature, for 
$W = 10^2 /k_c$ ($= 50$ nm) and a spatial separation between superconducting
contacts $L = 0.5 \; \mu$m (a), $1.5 \; \mu$m (b), and $2.5 \; \mu$m (c).}
\label{two}
\end{figure}

\begin{figure}
\begin{center}
\epsfxsize 5cm \epsfbox{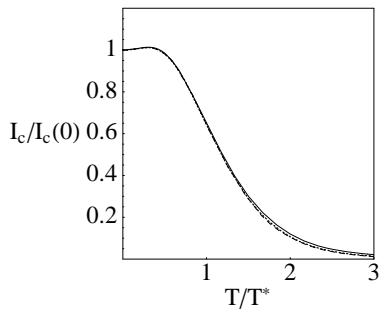}
\end{center}
\caption{Combined plot of $I_c^{(2D)} (T) / I_c^{(2D)} (0)$ represented as
a function of the scaled variable $T/T^* $ (with $T^* = v_F /2L $), where it is 
seen the collapse of the three curves corresponding to values of the distance 
$L = 0.5 \; \mu$m (full line), $1.5 \; \mu$m (dotted line), 
and $2.5 \; \mu$m (dashed line). }
\label{twop}
\end{figure}

In order to establish a comparison with experimental results, the relevant
behavior is given by the critical current represented as as function
of the temperature at fixed length $L$. The existence of a thermal length 
has a reflection here in the form of a crossover temperature $T^*$, which 
marks the strong decay of the critical current for $T > T^*$. We have
plotted in Fig. \ref{two} the critical current $I^{(2D)}_c (T)$, computed
from Eq. (\ref{2D}), at different values of $L$ between $0.5 \; \mu {\rm m}$ 
and $2.5 \; \mu {\rm m}$. 
The shapes of the curves in the figure are quite similar, and it can be checked 
that they can be collapsed into a single universal curve after rescaling
the temperature by $T^* \propto v_F / k_B L $, as shown in Fig. \ref{twop}. This
is consistent with the expression of the critical current in Eq. (\ref{2D}), 
where it it seen that the effect of a variation of the length $L$ on
$I_c^{(2D)} (T) / I_c^{(2D)} (0)$
can be compensated by a suitable change in the scale of $T$, in the regime 
where the critical current is not sensitive to the precise value of $k_c $.

We observe that the behavior of the critical 
current is in all cases quite stable for $T \ll T^*$ and that there is even 
an upturn before the abrupt drop at the crossover temperature. 
These features have been also found in the theoretical investigation of 
the supercurrents in one-dimensional (1D) electron systems\cite{faz} and 
in carbon nanotubes\cite{prl}. The shape of the critical
currents obtained there is qualitatively similar to that of 
the curves in Fig. \ref{two}. A major difference is however that the decay
of the supercurrents in the carbon nanotubes is given by a $1/L$ dependence
in the ballistic regime, instead of the much stronger power-law
decay (\ref{powerl}) in graphene.

It is worth mentioning at this point the experiment reported in Ref. 
\onlinecite{orsay}, in which the properties of a graphene Josephson junction 
have been measured in the regime of
large distance between superconducting electrodes. In the experimental
setup described there, the minimum distance between 
superconducting contacts can be estimated as $\approx 2.5 \; \mu {\rm m}$.
While no supercurrent was observed below the critical temperature $T_c$ of 
the electrodes ($\approx 4$ K), a signature of the proximity effect was 
obtained in the measurements of the resistance as a function of temperature,
in the form of a sharp decrease at $T \approx 1$ K. Quite remarkably, this
value of $T$ is in good correspondence with the crossover temperature that 
we find in our model for a distance $L = 2.5 \; \mu {\rm m}$, as can be
seen from Fig. \ref{two}(c). It is therefore likely that the sharp decrease
measured in the resistance has its origin in the same suppression of the 
thermal effects that enhances
the supercurrents at $T < T^*$. We also notice that the 
prediction from our model is that the critical currents for such a large
value of $L$ should be well below the scale of 1 nA. This may explain the
failure to establish a supercurrent in the experiment of 
Ref. \onlinecite{orsay}, and it may also anticipate better perspectives in
experiments with suitably short graphene junctions.

\section{Supercurrents at finite doping}

\begin{figure}[h]
\begin{center}
\epsfxsize 6cm \epsfbox{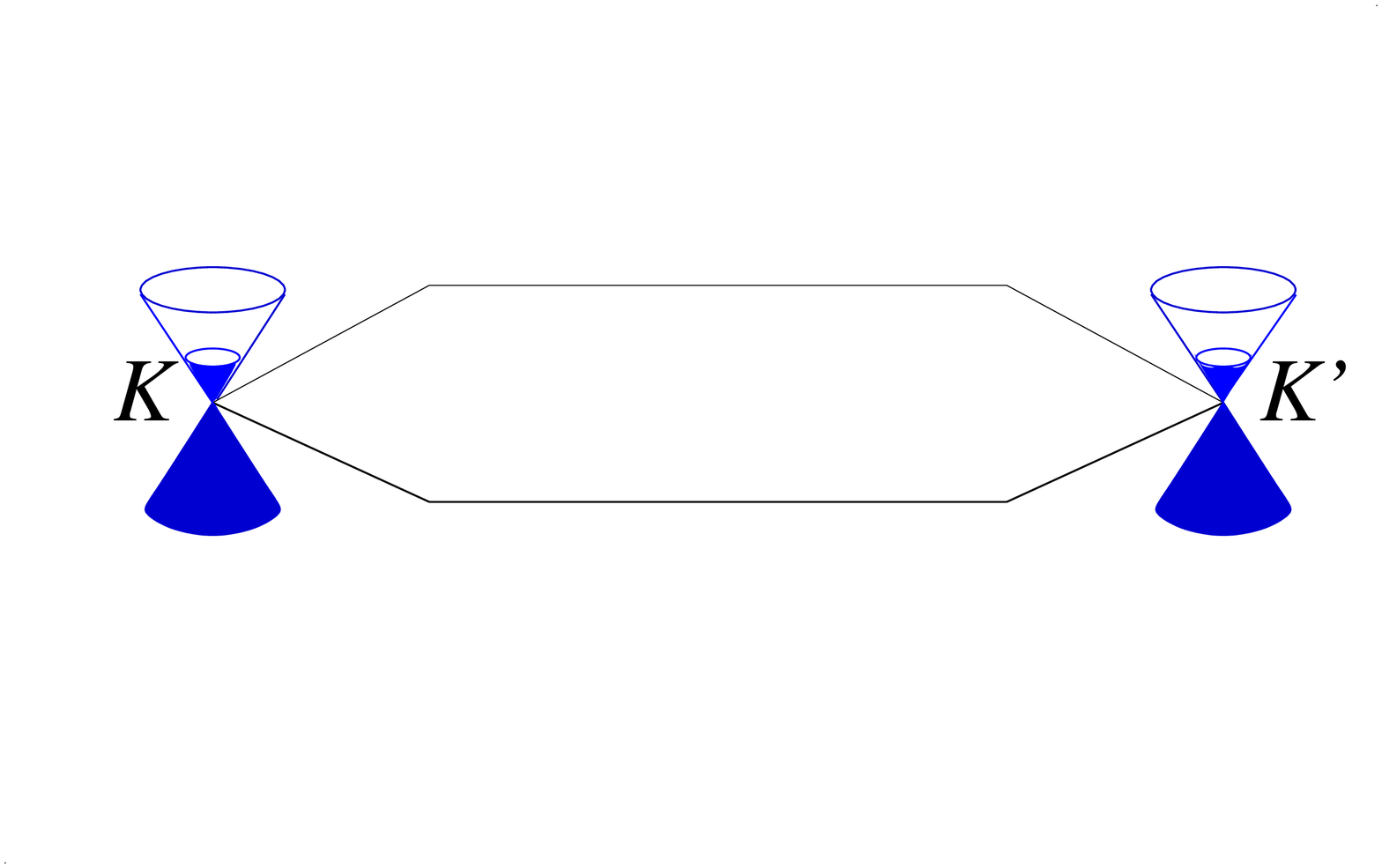}
\end{center}
\caption{Schematic representation of the two independent Dirac valleys at
the corners of the hexagonal Brillouin zone, showing the regions of occupied 
(dark) and unoccupied (white) energy levels in doped graphene.}
\label{hexadop}
\end{figure}

We have seen that the origin of the relative smallness of the critical 
currents in undoped graphene lies in the vanishing density of states 
at the Dirac point. Therefore, a straightforward way to enhance 
the supercurrents may simply consist in shifting the Fermi level away from 
the charge neutrality point, as shown in Fig. \ref{hexadop}. 
In practice, this can be achieved by 
doping the graphene sheet. In our theoretical framework,
we will assume that this effect can be accounted for by means of
a finite chemical potential $\mu $ . Thus, the hamiltonian for
the graphene part of the junction will now read:
\begin{equation}
H_0   =    \int d^2 r \; \Psi^{(a) \dagger}_{\sigma} (\mathbf{r}) \:
  \left( v_F \mbox{\boldmath $\sigma$}^{(a)}\cdot\mbox{\boldmath $\partial$} 
                   - \mu \right) \:     \Psi^{(a)}_{\sigma} (\mathbf{r})
\label{hamm}
\end{equation}

Working at $\mu \neq 0$ leads to significant modifications in the propagator
of the Dirac fermions and in the Cooper-pair propagator. The Dirac propagator 
corresponding to the hamiltonian (\ref{hamm}) turns out to be 
(for $\mu > 0$)\cite{chin}
\begin{eqnarray}
G^{(a)} (\mathbf{k}, \omega)  & =  &
  (  \omega + v_F \mbox{\boldmath $\sigma$}^{(a)} \cdot \mathbf{k} )
    [ \frac{1}{\omega^2 - v_F^2 \mathbf{k}^2 + i \epsilon}   
    + i \pi
       \frac{\delta (\omega - v_F |\mathbf{k}|)}{v_F |\mathbf{k}|}
         \theta (\mu - v_F |\mathbf{k}|)  ]
\label{diracmu}
\end{eqnarray}
As shown in the Appendix, the representation (\ref{diracmu}) is nothing 
but a compact form 
of expressing the propagation of quasiparticles with $v_F |\mathbf{k}| > \mu$ 
and quasiholes with $\pm v_F |\mathbf{k}| < \mu$, in the particular case of 
conical dispersion.

The propagator (\ref{diracmu}) is very convenient to carry out calculations
in the many-body theory and, in particular, it allows us to compute the 
dependence on $\mu$ of the Cooper-pair propagator as a correction to the 
expression (\ref{zerot}) at $\mu = 0$. In this procedure, we observe that 
the second term in the r.h.s. of (\ref{diracmu}) does not introduce any
integrals requiring regularization in the diagrammatics of the 
Dirac theory. By computing then the 
Cooper-pair propagator according to Eq. (\ref{freeprop}), we obtain\cite{prb}
\begin{eqnarray}
D (\mathbf{k}, 0)  & = &  - \frac{1}{2\pi v_F^2} \mu  \;\;\;\;\;
  \;\;\;  \;\;\; \;\;\;  \;\;\;
           {\rm if}  \;\;\;    v_F |\mathbf{k}| < 2 \mu    \nonumber    \\
        & = &  - \frac{1}{8 v_F} |\mathbf{k}|
    + \frac{1}{4\pi v_F} |\mathbf{k}|
        \arcsin \left( \frac{2\mu }{v_F |\mathbf{k}|} \right)  
            - \frac{1}{2\pi v_F^2} \mu  \;\;\;\;\;
                {\rm if}  \;\;\;   v_F |\mathbf{k}| > 2 \mu
\label{propmu}
\end{eqnarray}

\begin{figure}
\begin{center}
\mbox{\epsfxsize 5.5cm \epsfbox{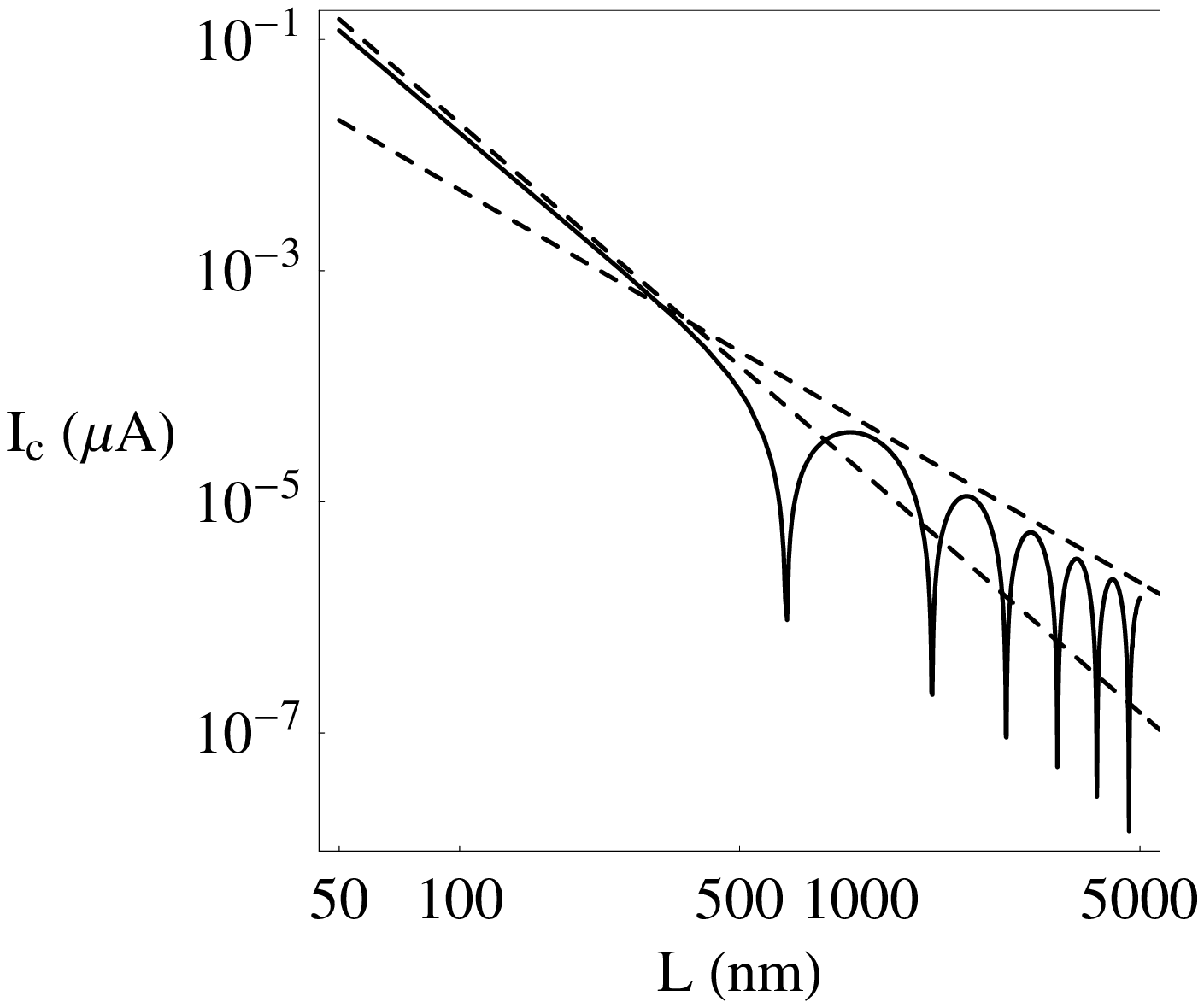} \hspace{0.45cm}
\epsfxsize 5.5cm \epsfbox{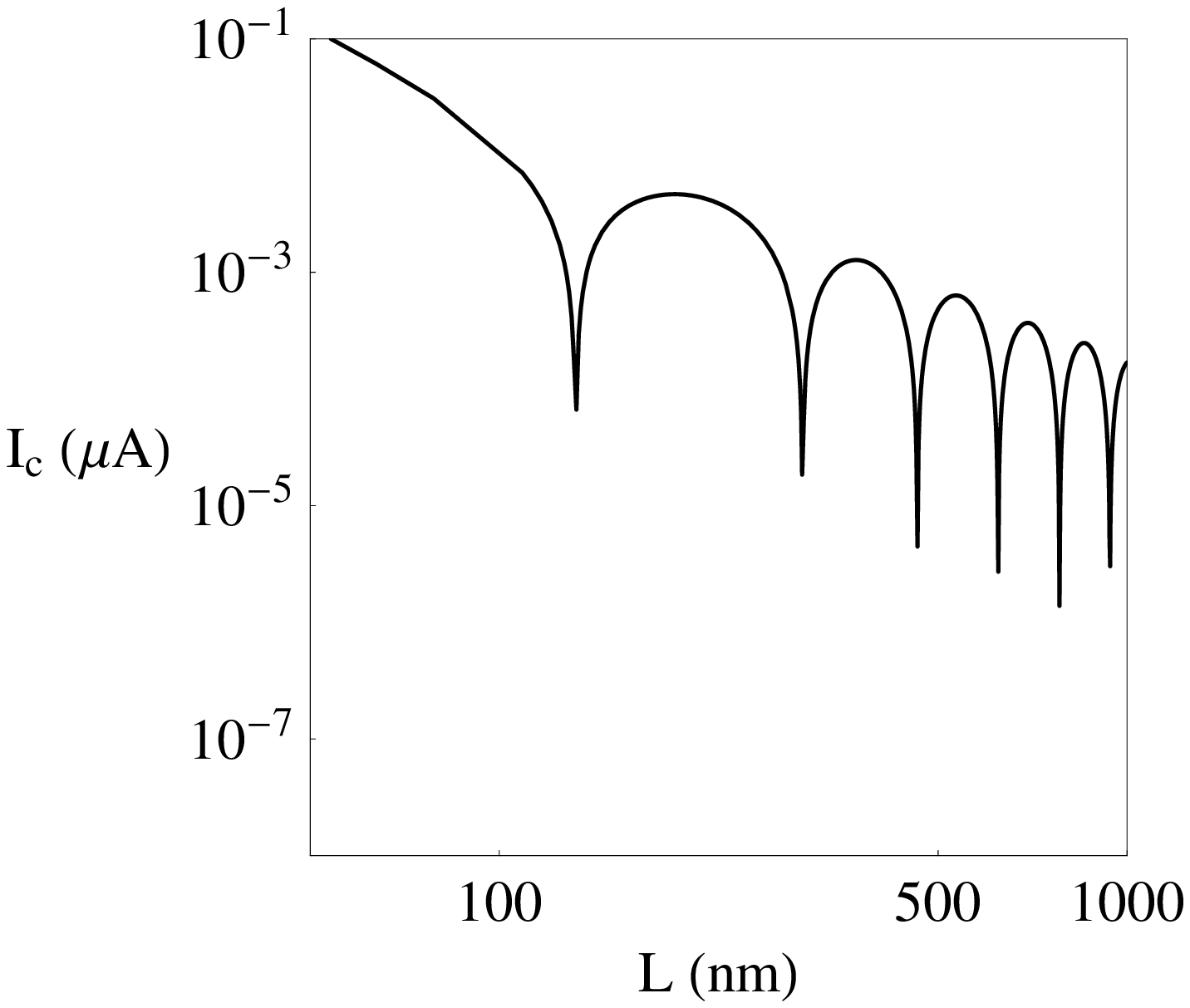}      \hspace{0.45cm}
\epsfxsize 5.5cm \epsfbox{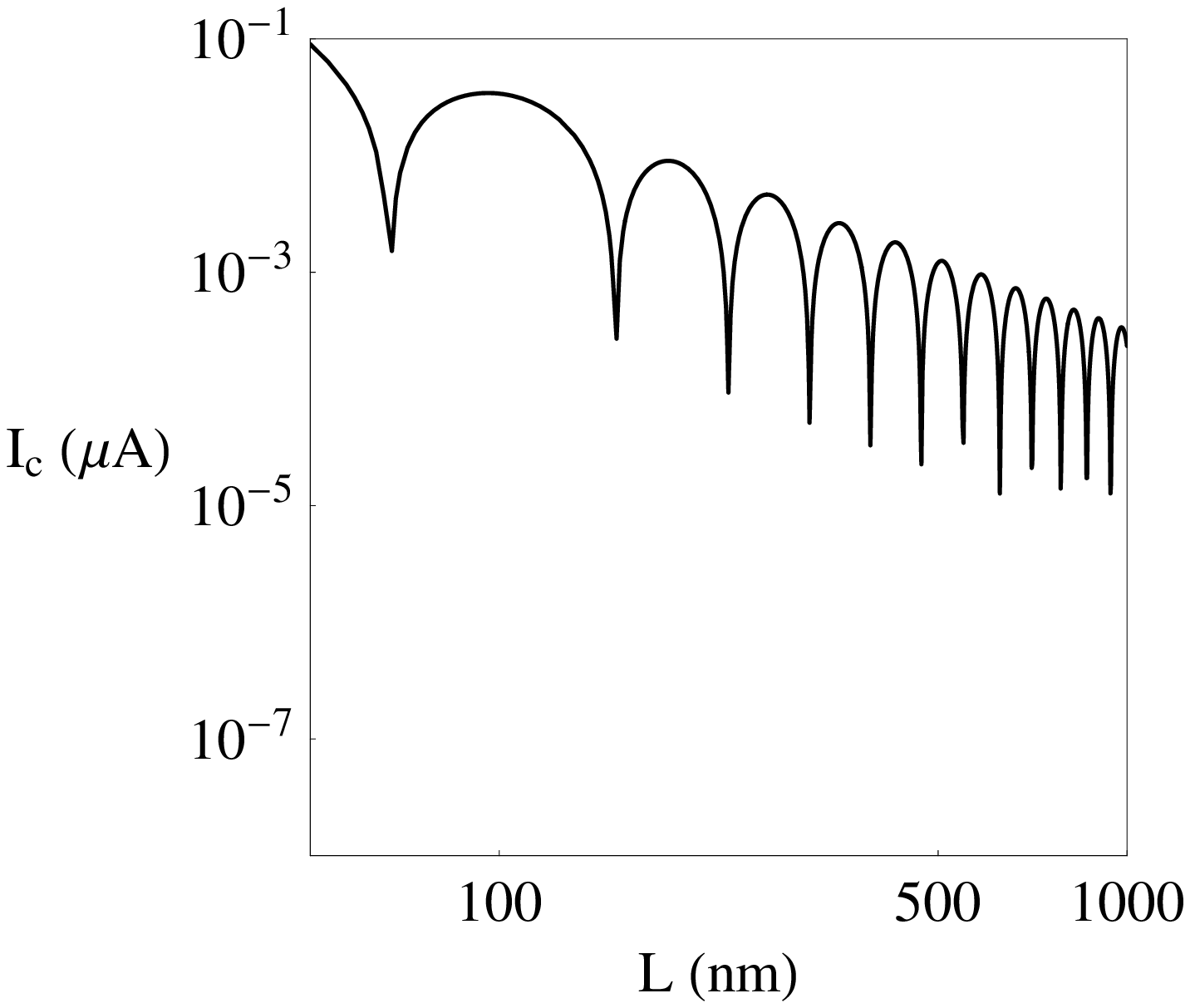}}\\
 \hspace{0.75cm}   (a) \hspace{5.45cm} (b)   \hspace{5.55cm} (c)
\end{center}
\caption{Plot of the zero-temperature critical current $I_c^{(2D)} (0)$  
as a function of the 
distance $L$, for $W = 10^2 /k_c$ ($= 50$ nm) and three different values of
the chemical potential $\mu = 1$ meV (a), 5 meV (b), and 10 meV (c). The 
dashed straight lines in Fig. 3(a) are drawn as a reference to the power-law
dependences $1/L^3$ and $1/L^2$.}
\label{three}
\end{figure}

At large values of $v_F |\mathbf{k}| \gg \mu $, we recover from  (\ref{propmu}) 
the linear dependence on the momentum that is characteristic of the Cooper-pair
propagator in the undoped system. However, the chemical potential introduces 
a clear deviation from that behavior at small $|\mathbf{k}| $, which has 
significant consequences in the decay of the supercurrent at long distances. 
This is illustrated in Fig. \ref{three}, where it can be appreciated the 
existence in general of a crossover length scale $L^*$ mediating the 
transition towards a softer power-law decay.

According to (\ref{propmu}), we can express the critical current 
$I_c^{(2D)} (0)$ at finite chemical potential in the form
\begin{equation}
I^{(2D)}_c (0) = I^{(2D)}_{c1} (0) + I^{(2D)}_{c2} (0)
\label{chp}
\end{equation}
with
\begin{eqnarray}
I^{(2D)}_{c1} (0)   &  = &  - \frac{1}{\pi }   e W^2           
         \mu \int_0^{\infty }  \frac{dk}{2\pi } \; |{\bf k}| \; 
        J_0 (|{\bf k}| L)  e^{- |{\bf k}| /k_c }           \\
I^{(2D)}_{c2} (0)    & =  &  -  \frac{1}{2 \pi } e v_F  W^2           
          \int_{2 \mu / v_F }^{\infty }  \frac{dk}{2\pi } \; |{\bf k}|^2 \; 
      \arccos \left( \frac{2\mu }{v_F |\mathbf{k}|} \right)  \;
        J_0 (|{\bf k}| L)  e^{- |{\bf k}| /k_c }
\end{eqnarray}
The first contribution to (\ref{chp}) is not relevant, since
we have 
\begin{equation}
I^{(2D)}_{c1} (0) =   - \frac{1}{\pi }  e W^2           
         \mu  \frac{k_c^2}{\sqrt{(k_c^2L^2 + 1)^3}}
\end{equation}
which is smaller than the estimate (\ref{powerl}) at $\mu = 0$ by a 
factor $\mu / v_F k_c $. The second contribution may change however the 
behavior of the critical current at large $L$, as the integrand is not 
analytic at $|{\bf k}| = 2\mu / v_F $. The expression for 
$I^{(2D)}_{c2} (0)$ is actually finite in the limit 
$k_c \rightarrow \infty $, and we obtain
\begin{eqnarray}
I^{(2D)}_{c2} (0)  & \sim &   - e v_F W^2 \frac{\mu^3 }{v_F^3}
    \int_{1}^{\infty }  dx \; x^2 \; 
      \arccos \left( \frac{1 }{x} \right)  \;
        J_0 ( (2 \mu  L/ v_F) x )                                    \\
       &  \sim  &  e W^2  \mu  \frac{1}{L^2}        \;\;\;\;\;\;\;\;\;\;
     \;\;\;\;\;           {\rm for}  \;\;\;   \mu L /v_F  \gg 1
\label{decmu}
\end{eqnarray}

We have to stress anyhow that $I^{(2D)}_{c2} (0)$ has oscillations
has a function of $L$, arising from the own behavior of the Bessel function
$J_0 $. The power-law decay (\ref{decmu}) applies then to the envelope of the 
maxima of the critical current, as it is illustrated in Fig. \ref{three}(a).
There it can be appreciated the crossover from the $1/L^3 $ behavior to the
oscillatory regime with softer power-law decay.
From the numerical results represented in the figure, it can be checked
that the $1/L^2 $ behavior is followed with great accuracy at larges values 
of $L$ (compared to $v_F /\mu $).

From a practical point of view, the most important result that we obtain is
the significant enhancement of the critical currents at moderate values of the
chemical potential. This is clearly observed in the plots of Fig. \ref{three},
where the crossover to the $1/L^2 $ decay is always found at a length scale 
consistent with the theoretical estimate $L^* \sim v_F /\mu $. For a chemical
potential $\mu \approx 10$ meV, for instance, that scale is 
$ \approx 50$ nm. The critical currents can be then enhanced to values 
above the nanoampere scale for spatial separation between superconducting
contacts $L \gtrsim 500$ nm  (assuming thin electrodes as in our case with 
$W \sim 50$ nm). This should open good perspectives to establish supercurrents
in graphene Josephson junctions by suitable doping of the samples.

\section{Conclusion}

In this paper we have adopted a framework suited to address the
many-body properties of graphene Josephson junctions. We have described
the development of the supercurrents through the tunneling and propagation
of Cooper pairs in the graphene part of the junction, with the aim of
investigating the dependence of the critical currents on relevant variables
like the distance between the superconducting contacts, the temperature,
and the doping level. We have been able then to characterize different 
regimes in the behavior of the supercurrents, depending on the relation 
between those variables. 

The supercurrents have a natural tendency to decay in the graphene part
of the Josephson junction, following in general a power-law behavior with
respect to the distance $L$ between the superconducting contacts. Such a 
power-law decay is particularly strong in undoped graphene, given the 
vanishing density of states at the charge neutrality point. We have shown
that the critical currents display then at zero temperature a $1/L^3 $
dependence on the distance $L$. When temperature effects are taken into 
account, there is always a finite thermal length $L_T $ (of the order of 
$\sim v_F / k_B T $) beyond which the supercurrents are further suppressed,
due to the disruption of the Cooper pairs by many-body effects. When this
takes place, the supercurrents enter a regime where the natural decay is
given by a $1/L^5 $ dependence. 

On the other hand, many-body effects can
be also used in our benefit to enhance the critical currents, in this case 
by shifting the Fermi level away from the charge neutrality point. This 
can be achieved in our framework by means of a chemical potential
$\mu \neq 0$. Inducing in this way a finite density of states at the Fermi
level, we have seen that the critical currents are enhanced beyond a new
crossover length $L^* \sim v_F / \mu $. This is actually the scale at
which the effects of the finite charge density start being felt, marking
the transition from the previously discussed $1/L^3 $ behavior to a 
softer $1/L^2 $ decay of the supercurrents at long distances.

At this point, it is interesting to note that the $1/L^3 $ 
decay at zero temperature in undoped graphene is similar to the behavior 
found in the investigation of mesoscopic junctions made of a diffusive 
metal\cite{dubos}. 
In this case, the product of the critical current times the normal resistance 
of the metal is proportional to the Thouless energy, which depends on
length $L$ as $1/L^2 $. This implies consequently a $1/L^3 $ decay
of the critical current, which we have seen is characteristic of graphene
under conditions of ballistic transport. The reminiscence of some of the
properties  of clean graphene with respect to the behavior of a disordered
normal metal has been remarked in several other instances\cite{twor,tit}. 
We have to point
out, however, that this resemblance does not go farther in our case, regarding
other regimes of the graphene Josephson junction. In particular, we have seen
that the critical current does not follow an exponential decay at distances 
larger than the thermal length. The decay of the critical current is always 
given in graphene by a power-law, which can be seen as a consequence of the 
perfect scaling of the low-energy Dirac theory.

We have also to stress that our results refer to Josephson junctions with
graphene layers which have large dimensions in both the longitudinal direction
along the junction and the transverse direction. This condition comes from 
our consideration of a system which is truly 2D, where in particular the 
size in the direction transverse to the junction is not constrained by
the width of the superconducting contacts. In this regard, the 
situation is quite different to the case of long but narrow junctions, where 
the small transverse dimension may lead to the quantization of the transverse
component of the momentum. 
In such circumstances, the behavior of the system may be rather dictated by a 
1D propagation of the Cooper pairs, which is known to lead to a 
$1/L $ decay of the supercurrents in the ballistic regime\cite{faz,prl}. 

Anyhow, the great advantage of the graphene Josephson junctions
is that the interaction effects have little significance at the temperatures 
required to measure the supercurrents. In the long 1D junctions made of carbon 
nanotubes, for instance, it is known that the Coulomb interaction may induce
a strong power-law suppression of the density of states, with the consequent 
reflection in the decay of the supercurrent\cite{prl}. In 2D graphene, however, 
the electron system has the tendency to become less correlated at low energies, 
with a strong renormalization of the Coulomb interaction that makes it 
practically irrelevant at the temperature scale of 1 K \cite{nos2,prbr}.

In conclusion, our results highlight the role of the different
parameters conforming the geometry of graphene Josephson junctions in the 
determination of the critical currents. We have seen that the interplay with
variables like the temperature and the doping level is what establishes the
different regimes of a junction. This information may be useful in the design
of experiments, for the purpose of enhancing the magnitude of the 
critical currents in real devices.

\section*{Acknowledgments}

We thank H. Bouchiat and F. Guinea for useful comments.
The financial support of the Ministerio de Educaci\'on y Ciencia
(Spain) through grant FIS2005-05478-C02-02 is gratefully
acknowledged. E.P. is also financially supported by CNISM (Italy).

\section*{Appendix}

\subsection{Lowest-order contribution to the critical current}

The Josephson current $I_s $ can be computed in a perturbative framework
by expanding the free energy in Eq. (\ref{joseph}) in powers of the tunneling 
amplitude $t$. The first nonvanishing contribution is found to fourth order 
in this expansion, as the statistical average of operators leads then to the 
appearance of the condensates of the two superconductors in the junction. 
We have actually
\begin{eqnarray}
I_s  & \approx &  2e \frac{\partial }{\partial \chi}
k_B T \; t^4 \prod_{i=1}^{4}  
    \int_0^{1/k_B T}  d \tau_i    \int_0^W dy_i   
           \langle \Psi_{S1,\uparrow} (0,y_1;-i\tau_1) 
     \Psi_{S1,\downarrow} (0,y_2;-i\tau_2)  \rangle       \nonumber     \\
  &  &    \times  \langle \Psi^{(a) \dagger}_{\uparrow} (0,y_1;-i\tau_1)
            \Psi^{(-a) \dagger}_{\downarrow} (0,y_2;-i\tau_2)  
             \Psi^{(b)}_{\uparrow} (L,y_3; -i\tau_3 )
       \Psi^{(-b)}_{\downarrow} (L,y_4; -i\tau_4 )   \rangle    \nonumber  \\
  &  &    \times   \langle \Psi^{\dagger}_{S2,\uparrow} (L,y_3;-i\tau_3) 
     \Psi^{\dagger}_{S2,\downarrow} (L,y_4;-i\tau_4)  \rangle
\label{pert}
\end{eqnarray}

We can apply to Eq. (\ref{pert}) the approximations pertinent
to the regime we want to study in the paper. Focusing on the case of a 
large junction where the distance $L$ is much larger than the superconducting
coherence length $\xi $, the use of Eq. (\ref{cond}) and translational 
invariance in the variable $\tau $ leads to a maximum supercurrent $I_c $
(critical current)
\begin{eqnarray}
I_c (T)   & \approx &  2e \rho^2  t^4 W^2  \prod_{i=1}^{2}  
     \int_0^W dy_i   \int_0^{1/k_B T}  d \tau   \:  
           \langle \Psi^{(a) \dagger}_{\uparrow} (0,y_1;0)
            \Psi^{(-a) \dagger}_{\downarrow} (0,y_1;0)   
                   \Psi^{(b)}_{\uparrow} (L,y_2; -i\tau )
          \Psi^{(-b)}_{\downarrow} (L,y_2; -i\tau )   \rangle 
\end{eqnarray}

At this point, it becomes
convenient to factor out the relative tunnel conductances at the contacts,
which are each given by the dimensionless quantity $\rho t^2 W /v_F $.
We concentrate then on the behavior of the critical current intrinsic 
to the 2D graphene system, represented by the expression 
\begin{eqnarray}
I_c^{(2D)} (T) & \approx &  2e v_F^2 \int_0^W dy_1 \int_0^W dy_2
     \int_0^{1/k_B T}  d \tau    \:                     
      \langle \Psi^{(a) \dagger}_{\uparrow} (0,y_1;0)
            \Psi^{(-a) \dagger}_{\downarrow} (0,y_1;0)   
              \Psi^{(b)}_{\uparrow} (L,y_2; -i\tau )
             \Psi^{(-b)}_{\downarrow} (L,y_2; -i\tau )   \rangle
\end{eqnarray}
This last equation highlights the connection between the critical current
and the propagator of the Cooper pairs, which plays a central role in the
discussion of Sections III and IV in the paper.

\subsection{Dirac propagator at $\mu \neq 0$}

In the many-body theory of Dirac fermions, it is usual to write the 
Dirac propagator at finite charge density in the form (assuming 
$\mu > 0$)
\begin{eqnarray}
G^{(a)} (\mathbf{k}, \omega)  & =  &
  (  \omega + v_F \mbox{\boldmath $\sigma$}^{(a)} \cdot \mathbf{k} )
    [ \frac{1}{\omega^2 - v_F^2 \mathbf{k}^2 + i \epsilon}   
    + i \pi
       \frac{\delta (\omega - v_F |\mathbf{k}|)}{v_F |\mathbf{k}|}
         \theta (\mu - v_F |\mathbf{k}|)  ]
\label{diracmuap}
\end{eqnarray}

If we specialize the expression (\ref{diracmuap}) to modes such 
that the eigenvalue $\varepsilon (\mathbf{k})$ of the matrix  
$v_F  \mbox{\boldmath $\sigma$}^{(a)} \cdot \mathbf{k}$ is positive,
we get 
\begin{eqnarray}
\left. G^{(a)} (\mathbf{k}, \omega)  
    \right|_{\varepsilon (\mathbf{k}) = v_F |\mathbf{k}|}  & = &   
  \frac{1}{\omega - v_F |\mathbf{k}|} 
   - i \pi \frac{\omega + v_F |\mathbf{k}|}{2v_F |\mathbf{k}|} 
              \delta (\omega - v_F |\mathbf{k}|)            \nonumber      \\
    & = &  \frac{1}{\omega - v_F |\mathbf{k}| + i \epsilon }
\end{eqnarray}
for $v_F |\mathbf{k}| > \mu$, and 
\begin{eqnarray}
\left. G^{(a)} (\mathbf{k}, \omega)  
    \right|_{\varepsilon (\mathbf{k}) = v_F |\mathbf{k}|}  & = &   
  \frac{1}{\omega - v_F |\mathbf{k}|} 
   + i \pi \frac{\omega + v_F |\mathbf{k}|}{2v_F |\mathbf{k}|} 
              \delta (\omega - v_F |\mathbf{k}|)            \nonumber      \\
    & = &  \frac{1}{\omega - v_F |\mathbf{k}| - i \epsilon }
\end{eqnarray}
for $v_F |\mathbf{k}| < \mu$. 
We observe that, in either case, the expression of $G^{(a)} (\mathbf{k}, \omega)$
agrees with the conventional propagation of a fermion, with the correct
$\pm i \epsilon $ prescription depending on whether it corresponds to a 
quasiparticle or a quasihole excitation.

On the other hand, we always have in the case of negative eigenvalue 
$\varepsilon (\mathbf{k})$
\begin{eqnarray}
\left. G^{(a)} (\mathbf{k}, \omega)  
    \right|_{\varepsilon (\mathbf{k}) = - v_F |\mathbf{k}|}  & = &   
  \frac{1}{\omega + v_F |\mathbf{k}|} 
   - i \pi \frac{\omega - v_F |\mathbf{k}|}{2v_F |\mathbf{k}|} 
              \delta (\omega + v_F |\mathbf{k}|)            \nonumber      \\
    & = &  \frac{1}{\omega + v_F |\mathbf{k}| - i \epsilon }
\end{eqnarray}
which is also in agreement with the expected propagation for a quasihole in
the valence band of graphene.

\end{document}